\documentclass{article}
\usepackage{cite}
\usepackage{graphicx}
\usepackage{dcolumn}

\begin{document}

\date{}
\title{Extremely broken generalized $\mathcal{PT}$ symmetry}
\author{Francisco M. Fern\'{a}ndez\thanks{%
fernande@quimica.unlp.edu.ar} \\
INIFTA, DQT, Sucursal 4, C.C 16, \\
1900 La Plata, Argentina}
\maketitle

\begin{abstract}
We discuss some simple H\"uckel-like matrix representations of non-Hermitian
operators with antiunitary symmetries that include generalized $\mathcal{PT}$
(parity transformation followed by time-reversal) symmetry. One of them
exhibits extremely broken antiunitary symmetry (complex eigenvalues for all
nontrivial values of the model parameter) because of the degeneracy of the
operator in the Hermitian limit. These examples illustrate the effect of
point-group symmetry on the spectrum of the non-Hermitian operators. We
construct the necessary unitary matrices by means of simple graphical
representations of the non-Hermitian operators.
\end{abstract}

\section{Introduction}

\label{sec:intro}

Since the seminal paper by Bender and Boettcher\cite{BB98} on parity-time
symmetry in non-relativistic quantum mechanics, there has been great
interest in all kinds of physical problems that lead to $\mathcal{PT}$%
-symmetric equations (see\cite{B07} and references therein). Strictly
speaking, the parity operation is given by $\mathcal{P}f(\mathbf{x})=f(-%
\mathbf{x})$, where $\mathbf{x}$ represents the set of coordinates that
describe the system. However, in some applications other operators $\mathcal{%
P}$ that satisfy $\mathcal{P}^{2}=\hat{1}$, where $\hat{1}$ is the identity
operator, have served as parity, such as, for example, the $2\times 2$
matrix representation chosen by Bender\cite{B07}. In this way, several fancy
names emerged, like partial $\mathcal{PT}$ symmetry\cite{BKB15} or
reverse-order partial $\mathcal{PT}$ symmetry\cite{F15}. Such kind of
pseudo-parity operators are particular cases of the unitary operators that
lead to antiunitary symmetries\cite{F15,W60,F14}.

Most studies based on parameter-dependent operators are commonly restricted
to the range of model parameter values for which the eigenvalues are real or
appear as complex-conjugate pairs, which gives rise to the so-called
unbroken (all the eigenvalues are real) or broken (one or more eigenvalues
are complex) $\mathcal{PT}$ symmetry\cite{B07} (and references therein). If $%
\lambda $ is the model parameter then $\mathcal{PT}$ symmetry is unbroken
for $0\leq \lambda <\lambda _{c}$ and broken for $\lambda >\lambda _{c}$,
where, typically, the Hamiltonian operator is Hermitian for $\lambda =0$. On
the other hand, some authors have searched for $\mathcal{PT}$-symmetric
non-Hermitian operators with complex eigenvalues for all non-trivial values
of the model parameter\cite{FG14b,FG14c,FG15}. It was argued that the
occurrence of this kind of \textit{extremely broken} $\mathcal{PT}$
symmetry, where, for example, $\lambda _{c}=0$, is due to the point-group or
dynamical symmetry of the physical problem and can be predicted by
first-order perturbation theory\cite{FG14b,FG14c,FG15}.

There has recently been interest in $\mathcal{PT}$-symmetry in molecular
systems. Burton et al\cite{BTL19} derived the conditions for $\mathcal{PT}$%
-symmetry in the context of electronic structure theory within the
Hartree-Fock (HF) approximation. These authors showed that the HF orbitals
are symmetric with respect to the $\mathcal{PT}$ operator if and only if the
effective Fock Hamiltonian is $\mathcal{PT}$-symmetric.

The purpose of this paper is to illustrate the concept of antiunitary
symmetry\cite{W60} as a generalization of $\mathcal{PT}$ symmetry, as well
as the effect of point-group symmetry\cite{T64,C90} on the occurrence of
extremely broken antiunitary symmetry. As illustrative examples we choose
non-Hermitian versions of the H\"{u}ckel model proposed many years ago for
the treatment of conjugated molecules\cite{P68,AD02} and also H$_{n}$
structures\cite{AD02}.

In section~\ref{sec:antiunit_sym} we summarize the main ideas about
antiunitary symmetry; in section~\ref{sec:example} we illustrate the effect
of point-group symmetry by means of two extremely simple examples based on
the H\"{u}ckel model; finally, in section~\ref{sec:conclusions} we summarize
the main results and draw conclusions.

\section{Antiunitary symmetry}

\label{sec:antiunit_sym}

The concept of antiunitary symmetry was developed by Wigner\cite{W60} some
time ago and later invoked in some discussions on generalized $\mathcal{PT}$
symmetry\cite{F15,F14,BBM02}. In this section we introduce it in a way
somewhat different from our earlier papers\cite{F15,F14}.

We begin the present discussion with the complex conjugation operator $K$
defined by
\begin{equation}
K\left| f\right\rangle =\left| f\right\rangle ^{*},  \label{eq:Kf}
\end{equation}
where $\left| f\right\rangle $ belongs to the physical vector space and $*$
stands for complex conjugation. Obviously, $K^{2}=\hat{1}$. From the well
known properties of complex conjugation we have
\begin{eqnarray}
K\left( \left| f\right\rangle +\left| g\right\rangle \right) &=&K\left|
f\right\rangle +K\left| g\right\rangle ,  \nonumber \\
Kc\left| f\right\rangle &=&c^{*}K\left| f\right\rangle ,
\label{eq:K-antilinear}
\end{eqnarray}
where $\left| g\right\rangle $ also belongs to the physical vector space and
$c$ is a complex number.

If $H$ is a non-Hermitian operator, then it follows from $KHK\left|
f\right\rangle =KH\left| f\right\rangle ^{*}=H^{*}\left| f\right\rangle $
that
\begin{equation}
KHK=H^{*}.  \label{eq:KHK}
\end{equation}
Suppose that there is a unitary operator $U$ ($U^{\dagger }=U^{-1}$, $%
U^{\dagger }$ being the adjoint of $U$) that satisfies
\begin{equation}
U^{\dagger }HU=H^{*}=KHK.  \label{eq:UHU_operator}
\end{equation}
Therefore,
\begin{equation}
A^{-1}HA=H,  \label{eq:AHA}
\end{equation}
where $A=UK$ is said to be an antiunitary symmetry of the physical system
represented by $H$. Equation (\ref{eq:AHA}) is equivalent to
\begin{equation}
\left[ H,A\right] =HA-AH=0.  \label{eq:[H,A]=0}
\end{equation}
It is worth noticing that $A$ satisfies equations (\ref{eq:K-antilinear}).

In some cases, there is a set of unitary operators $S_{U}=\left\{
U_{j},\;j=1,2,\ldots ,N\right\} $ that satisfy equation (\ref
{eq:UHU_operator}) and we can, therefore, construct a set of antiunitary
operators $S_{A}=\left\{ A_{j}=U_{j}K,\;j=1,2,\ldots ,N\right\} $ that leave
$H$ invariant as in equation (\ref{eq:AHA}). One can easily verify that $%
U_{i}U_{j}\notin S_{U}$ because $\left( U_{i}U_{j}\right) ^{\dagger
}HU_{i}U_{j}=H$. On the other hand, $A_{i}A_{j}\in S_{A}$.

If $\left| \psi \right\rangle $ is an eigenvector of $H$ with eigenvalue $E$%
\begin{equation}
H\left| \psi \right\rangle =E\left| \psi \right\rangle ,
\label{eq:H.psi=E.psi}
\end{equation}
then
\begin{eqnarray}
\left[ H,A\right] \left| \psi \right\rangle &=&HA\left| \psi \right\rangle
-AH\left| \psi \right\rangle =HA\left| \psi \right\rangle -AE\left| \psi
\right\rangle  \nonumber \\
&=&HA\left| \psi \right\rangle -E^{*}A\left| \psi \right\rangle =0.
\label{eq:[H,A].psi}
\end{eqnarray}
In other words: if $E$ is an eigenvalue of $H$ with eigenvector $\left| \psi
\right\rangle $ then $E^{*}$ is an eigenvalue of $H$ with eigenvector $%
A\left| \psi \right\rangle $.

If
\begin{equation}
A\left| \psi \right\rangle =a\left| \psi \right\rangle ,  \label{eq:A.psi}
\end{equation}
where $a$ is a complex number, for all the eigenvectors of $H$, we say that
the antiunitary symmetry is unbroken because it follows from equation (\ref
{eq:[H,A].psi}) that $H\left| \psi \right\rangle =E^{*}\left| \psi
\right\rangle $ and $E=E^{*}$. If, on the other hand, $\left| \psi
\right\rangle $ and $A\left| \psi \right\rangle $ are linearly independent
(for at least one eigenvector of $H$) , then $E^{*}$ is an eigenvalue $H$
with eigenvector $A\left| \psi \right\rangle $ and we say that the $\mathcal{%
PT}$ symmetry is broken.

In order to identify the antiunitary symmetries of a given operator $H$ we
just look for unitary operators $U$ that satisfy equation (\ref
{eq:UHU_operator}). In the following section we illustrate this point by
means of some simple examples.

\section{Simple examples}

\label{sec:example}

\begin{figure}[tbp]
\begin{center}
\includegraphics[width=9cm]{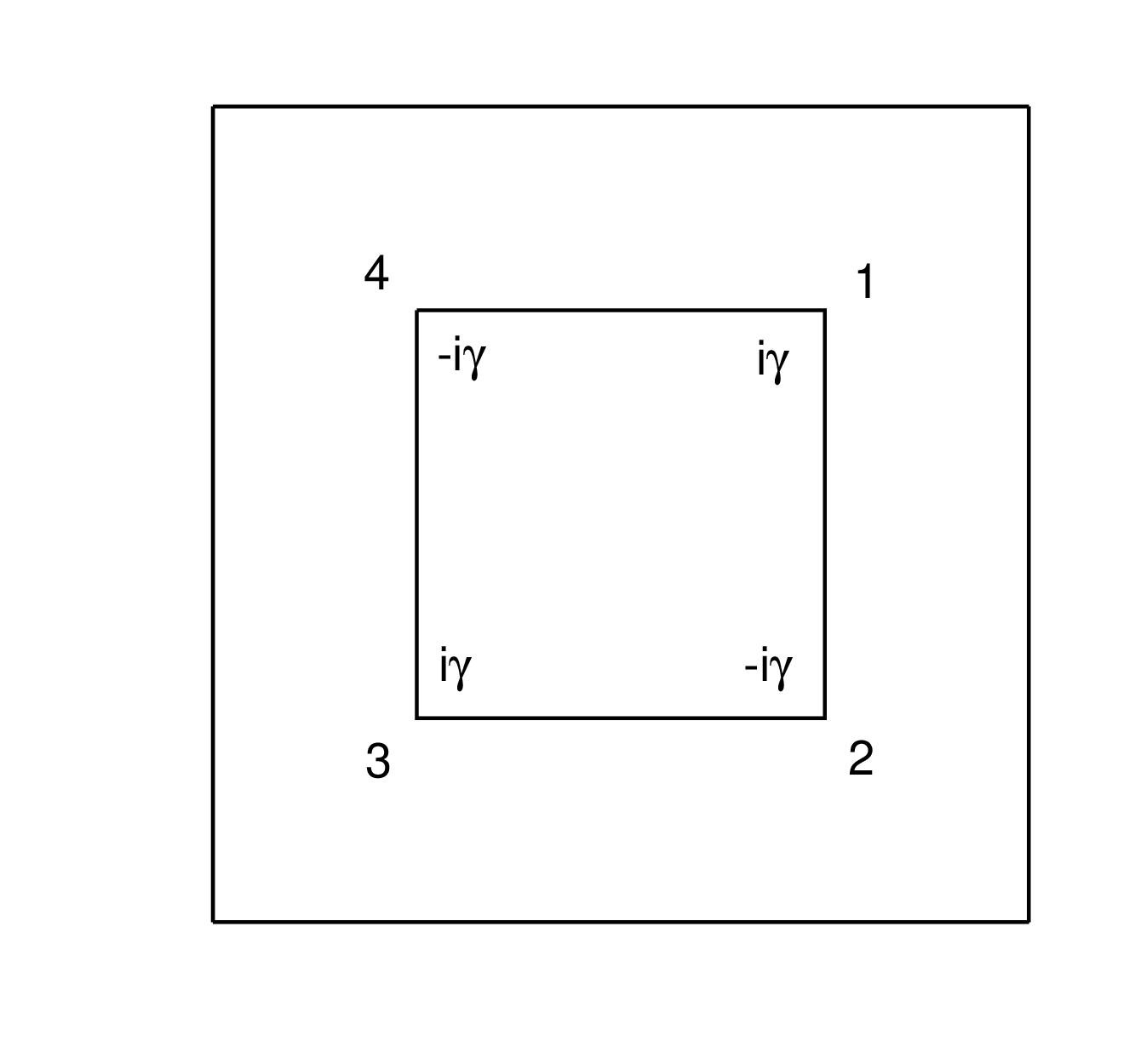}
\end{center}
\caption{Simple graphical representation of the periodic chain related to
the non-Hermitian operator (\ref{eq:H_4x4_per})}
\label{Fig:Cuad}
\end{figure}

In this section, we consider simple matrix representations of non-Hermitian
operators. Our first illustrative example is given by the matrix
\begin{equation}
\mathbf{H}_{P}(\gamma )=\left(
\begin{array}{llll}
i\gamma & 1 & 0 & 1 \\
1 & -i\gamma & 1 & 0 \\
0 & 1 & i\gamma & 1 \\
1 & 0 & 1 & -i\gamma
\end{array}
\right) ,  \label{eq:H_4x4_per}
\end{equation}
where $\gamma $ is real, that may be related to a H\"{u}ckel model for a
cyclic molecule\cite{P68,AD02} with complex diagonal elements. To facilitate
the construction of unitary matrices that satisfy (see the general equation (%
\ref{eq:UHU_operator}))
\begin{equation}
\mathbf{U}^{\dagger }\mathbf{HU}=\mathbf{H}^{*},  \label{eq:UHU}
\end{equation}
we resort to the graphical representation of $\mathbf{H}_{P}$ shown in
figure~\ref{Fig:Cuad}. It is worth mentioning that we will refer to unitary
matrices for the sake of generality, but in fact all the matrices discussed
in what follows are real and, consequently, orthogonal.

The unitary matrix
\begin{equation}
\mathbf{U}_{1}=\left(
\begin{array}{llll}
0 & 1 & 0 & 0 \\
1 & 0 & 0 & 0 \\
0 & 0 & 0 & 1 \\
0 & 0 & 1 & 0
\end{array}
\right) ,  \label{eq:U_1}
\end{equation}
that produces the transformation
\begin{equation}
\mathbf{U}_{1}\left(
\begin{array}{l}
c_{1} \\
c_{2} \\
c_{3} \\
c_{4}
\end{array}
\right) =\left(
\begin{array}{l}
c_{2} \\
c_{1} \\
c_{4} \\
c_{3}
\end{array}
\right) ,  \label{eq:U_1C}
\end{equation}
already satisfies equation (\ref{eq:UHU}). Since this matrix satisfies $%
\mathbf{U}_{1}^{2}=\mathbf{I}$, where $\mathbf{I}$ is the $4\times 4$
identity matrix, it is a good candidate for generalized parity $\mathcal{P}=%
\mathbf{P}=\mathbf{U}_{1}$. In order to realize how to construct the unitary
matrices in a straightforward way, note that this one is just a
representation of a rotation of $\pi $ radians about an axis through the
middle of the sides $1-2$ and $3-4$ of the square in figure~\ref{Fig:Cuad}.

Another candidate for generalized parity is
\begin{equation}
\mathbf{U}_{2}=\left(
\begin{array}{llll}
0 & 0 & 0 & 1 \\
0 & 0 & 1 & 0 \\
0 & 1 & 0 & 0 \\
1 & 0 & 0 & 0
\end{array}
\right) ,  \label{eq:U_2}
\end{equation}
that leads to
\begin{equation}
\mathbf{U}_{2}\left(
\begin{array}{l}
c_{1} \\
c_{2} \\
c_{3} \\
c_{4}
\end{array}
\right) =\left(
\begin{array}{l}
c_{4} \\
c_{3} \\
c_{2} \\
c_{1}
\end{array}
\right) .  \label{eq:U_2C}
\end{equation}
Note that $\mathbf{U}_{2}^{2}=\mathbf{I}$ because it is a rotation of $\pi $
about an axis through the middle of the square sides $1-4$ and $2-3$. The
unitary matrices just derived are not the only ones that satisfy equation (%
\ref{eq:UHU}).

The unitary matrix
\begin{equation}
\mathbf{U}_{3}=\left(
\begin{array}{llll}
0 & 1 & 0 & 0 \\
0 & 0 & 1 & 0 \\
0 & 0 & 0 & 1 \\
1 & 0 & 0 & 0
\end{array}
\right) ,  \label{eq:U_3}
\end{equation}
gives rise to the cyclic permutation
\begin{equation}
\mathbf{U}_{3}\left(
\begin{array}{l}
c_{1} \\
c_{2} \\
c_{3} \\
c_{4}
\end{array}
\right) =\left(
\begin{array}{l}
c_{2} \\
c_{3} \\
c_{4} \\
c_{1}
\end{array}
\right) .  \label{eq:U_3C}
\end{equation}
It is a representation of a rotation of $\pi /2$ about an axis through the
center of the square in figure~\ref{Fig:Cuad}. For this reason, $\mathbf{U}%
_{3}^{4}=\mathbf{I}$ which gives rise to two more matrices that satisfy
equation (\ref{eq:UHU}): $\mathbf{U}_{4}=\mathbf{U}_{3}^{2}$ and $\mathbf{U}%
_{5}=\mathbf{U}_{3}^{3}$. It is clear that $\mathbf{U}_{3}^{\dagger }\mathbf{%
U}_{1}\mathbf{U}_{3}=\mathbf{U}_{2}$ because the axes associated to $\mathbf{%
U}_{1}$ and $\mathbf{U}_{2}$ are perpendicular. Although $\mathbf{U}_{4}^{2}=%
\mathbf{I}$, it is not a suitable candidate for generalized $\mathcal{PT}$
symmetry because $\mathbf{U}_{4}^{-1}\mathbf{HU}_{4}=\mathbf{H}$.

The existence of more than one candidate for generalized parity is a good
reason for speaking of antiunitary symmetry instead of $\mathcal{PT}$
symmetry or generalized $\mathcal{PT}$ symmetry\cite{BBM02}, thus avoiding
fancy names like partial $\mathcal{PT}$ symmetry\cite{BKB15} or
reverse-order partial $\mathcal{PT}$ symmerty\cite{F15}, even when the
former was observed experimentally\cite{XHHBJHZJ22}.

There are two matrices that satisfy $\mathbf{U}_{j}^{-1}\mathbf{HU}_{j}=%
\mathbf{H}$, $j=6,7$, which represent unitary symmetries, and are unsuitable
for the construction of antiunitary operators. They are the matrix
representations of the rotations of $\pi $ about axes through opposite
vertices of the square in figure~\ref{Fig:Cuad}.

\begin{figure}[tbp]
\begin{center}
\includegraphics[width=9cm]{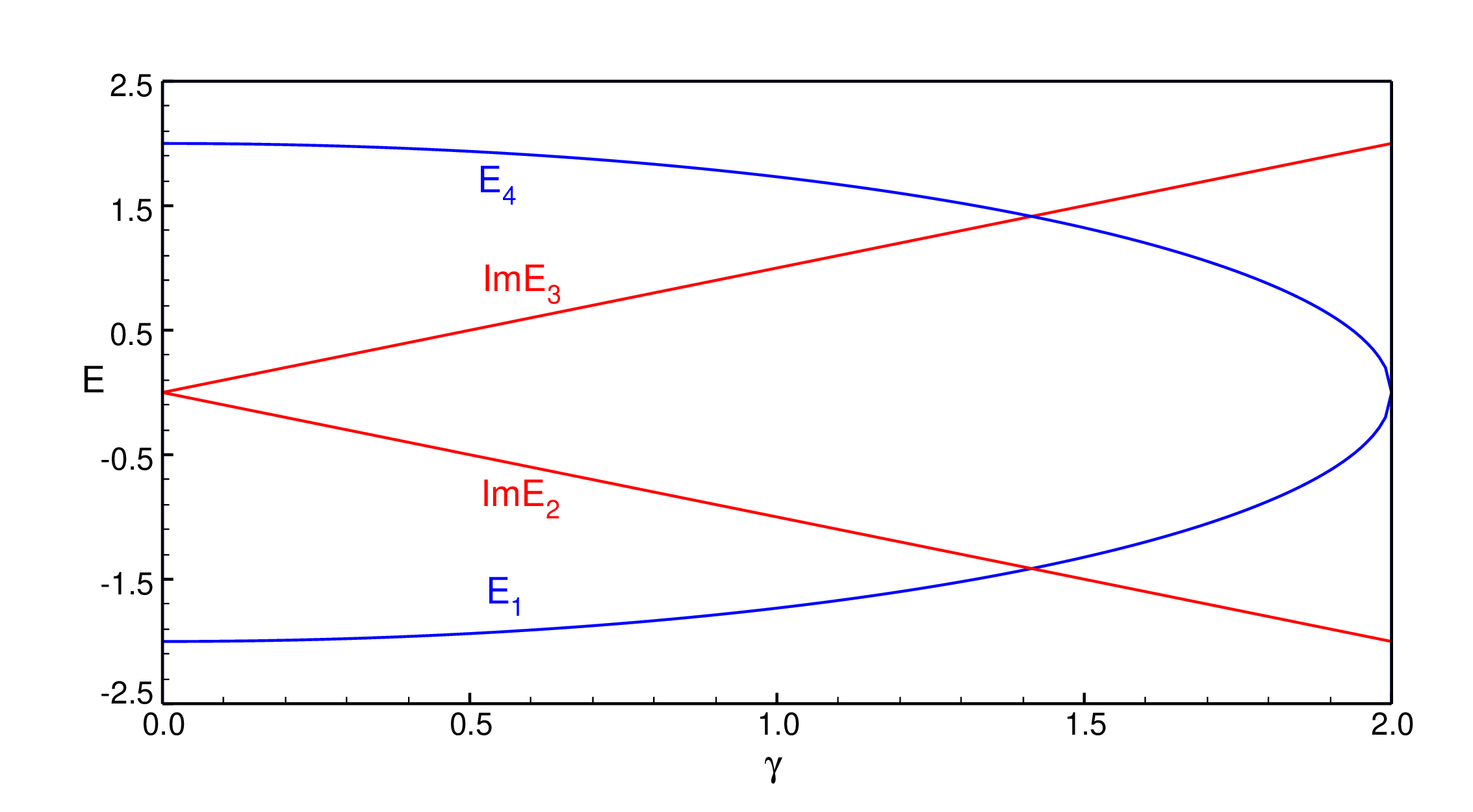}
\end{center}
\caption{Eigenvalues of the non-Hermitian operator (\ref{eq:H_4x4_per})}
\label{Fig:Eper}
\end{figure}

The eigenvalues of the matrix (\ref{eq:H_4x4_per}) are
\begin{equation}
E_{1}(\gamma )=-E_{4}(\gamma )=-\sqrt{4-\gamma ^{2}},\;E_{2}(\gamma
)=-E_{3}(\gamma )=-i\gamma ,  \label{eq:E_j_per}
\end{equation}
that show that the antiunitary symmetries constructed above are broken for
all $\gamma \neq 0$ (which we have chosen to call extremely broken $\mathcal{%
PT}$ symmetry). The reason is the degeneracy of $\mathbf{H}_{P}(0)$ as
discussed elsewhere for other non-Hermitian models\cite{FG14b,FG14c,FG15}.
Note that perturbation theory yields a $\gamma ^{2}$-power series for $E_{1}$
and $E_{4}$ and a $\gamma $-power series (only one term) for $E_{2}$ and $%
E_{3}$. Once again, we appreciate that degeneracy is the cause of the
extremely broken antiunitary symmetry. The degeneracy when $\gamma =0$ is
due to the point-group symmetry of $\mathbf{H}_{P}(0)$ as this matrix is
invariant under all the transformations $\mathbf{U}_{j}^{\dagger }\mathbf{H}%
_{P}(0)\mathbf{U}_{j}=\mathbf{H}_{P}(0)$ given by $G_{8}=\left\{ \mathbf{I},%
\mathbf{U}_{j},\;j=1,2,\ldots ,7\right\} $. This set of $8$ matrices is a
group isomorphic to $C_{4v}$ that includes a two-dimensional irreducible
representation $E$\cite{T64,C90}. Figure~\ref{Fig:Eper} shows the
eigenvalues of $\mathbf{H}_{P}(\gamma )$ for some values of the model
parameter. It is clear that the two complex eigenvalues stem from the
degenerate ones $E_{2}(0)=E_{3}(0)$ as argued in previous papers about other
non-Hermitian models\cite{FG14b,FG14c,FG15}.

\begin{figure}[tbp]
\begin{center}
\includegraphics[width=9cm]{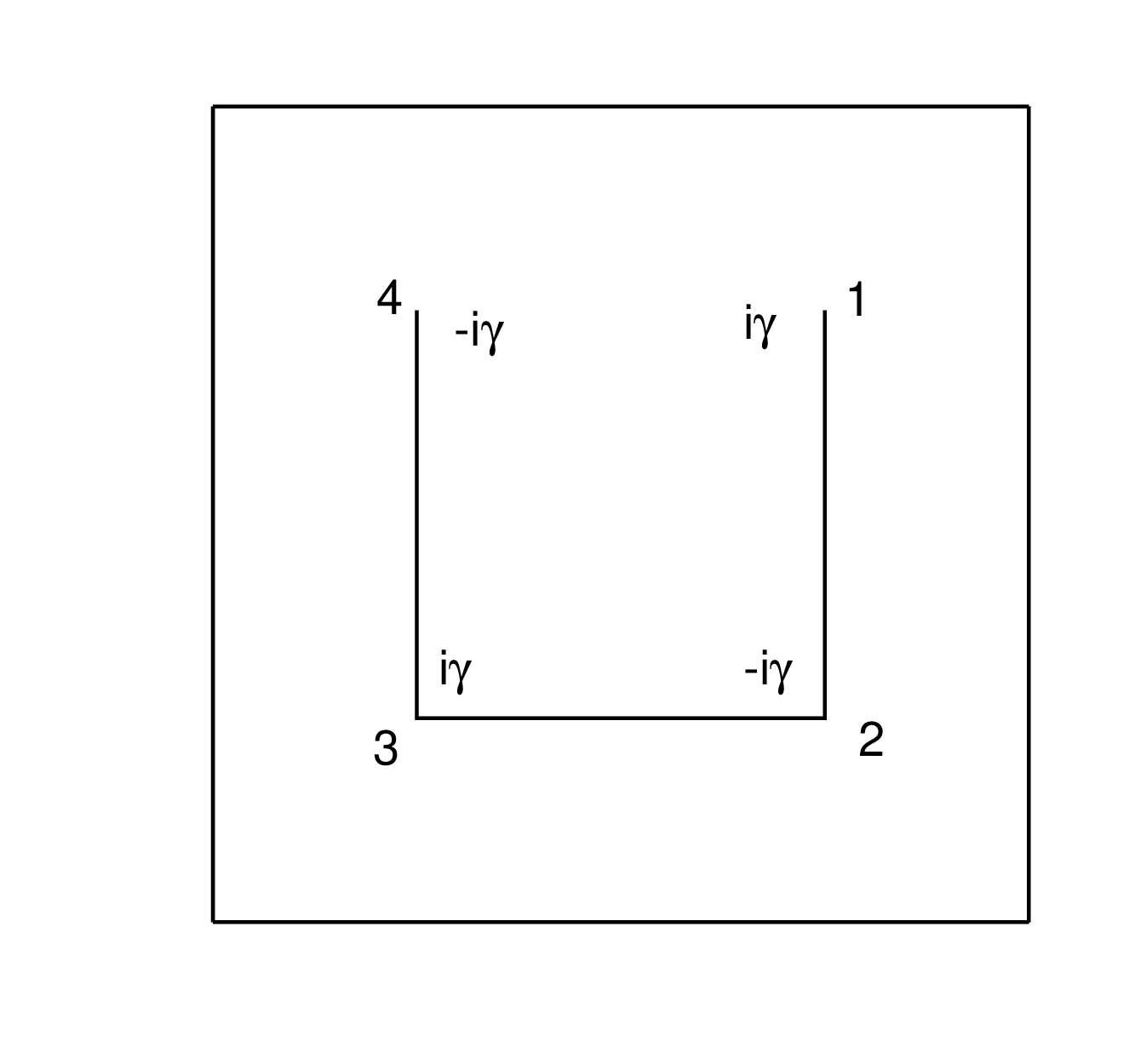}
\end{center}
\caption{Simple graphical representation of the open chain related to the
non-Hermitian operator (\ref{eq:H_4x4_open})}
\label{Fig:Open}
\end{figure}

Earlier discussions of the effect of point-group symmetry on antiunitary
symmetry\cite{FG14b,FG14c,FG15} suggest that a model with less symmetry than
(\ref{eq:H_4x4_per}) may exhibit real eigenvalues. One such model is given
by the H\"{u}ckel-like open chain in figure~\ref{Fig:Open} that represents
the matrix
\begin{equation}
\mathbf{H}_{O}(\gamma )=\left(
\begin{array}{llll}
i\gamma & 1 & 0 & 0 \\
1 & -i\gamma & 1 & 0 \\
0 & 1 & i\gamma & 1 \\
0 & 0 & 1 & -i\gamma
\end{array}
\right) .  \label{eq:H_4x4_open}
\end{equation}
In this case, the unitary matrix (\ref{eq:U_2}) is the sole candidate for an
antiunitary symmetry because it is the only one that satisfies equation (\ref
{eq:UHU}), i.e. $\mathbf{U}_{2}^{\dagger }\mathbf{H}_{O}\mathbf{U}_{2}=%
\mathbf{H}_{O}^{*}$. As indicated above, $\mathbf{U}_{2}$ is also a
candidate for generalized parity.

\begin{figure}[tbp]
\begin{center}
\includegraphics[width=9cm]{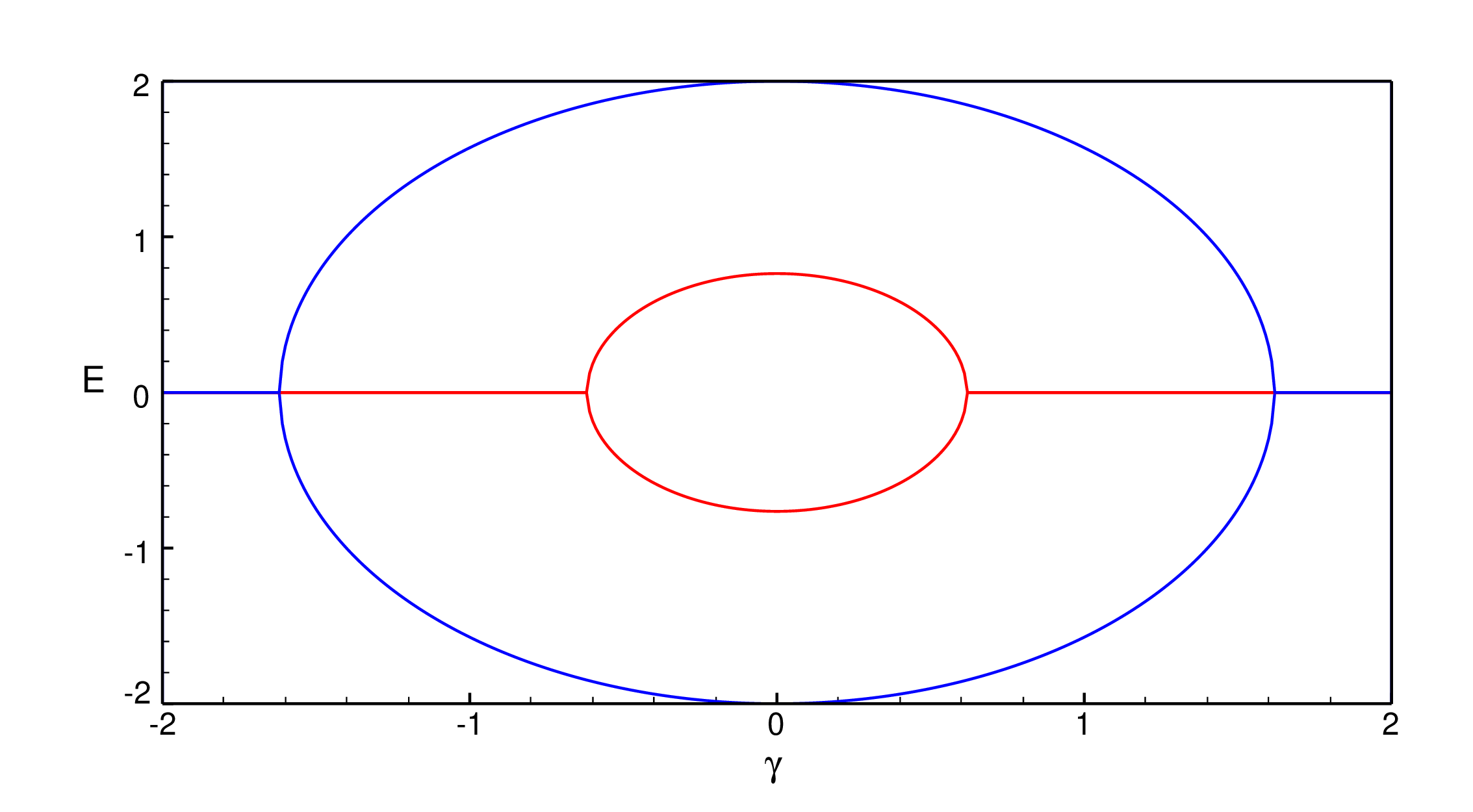}
\end{center}
\caption{Eigenvalues of the non-Hermitian operator (\ref{eq:H_4x4_open})}
\label{Fig:Eopen}
\end{figure}

The eigenvalues of $\mathbf{H}_{O}$ are
\begin{equation}
E_{1}=-E_{4}=-\frac{1}{\sqrt{2}}\sqrt{-2\gamma ^{2}+\sqrt{5}+3}%
,\;E_{2}=-E_{3}=-\frac{1}{\sqrt{2}}\sqrt{-2\gamma ^{2}-\sqrt{5}+3}.
\label{eq:E_j_open}
\end{equation}
There are four exceptional points $\gamma _{EP}=\pm \left( \sqrt{5}-1\right)
/2,\,\pm \left( \sqrt{5}+1\right) /2$ and, consequently, all the eigenvalues
are real when $|\gamma |<\left( \sqrt{5}-1\right) /2\approx 0.618$. They are
shown in figure~\ref{Fig:Eopen} that confirms the argument put forward
above. Note that $\mathbf{H}_{O}(0)$ does not exhibit degenerate states
because of its lower symmetry $C_{2}$ (the point-group being $G_{2}=\left\{
\mathbf{I},\mathbf{U}_{2}\right\} $) with two one-dimensional irreducible
representations\cite{T64,C90}.

The operator
\begin{equation}
\mathbf{H}_{O2}(\gamma )=\left(
\begin{array}{llll}
i\gamma  & 0 & 1 & 0 \\
0 & i\gamma  & 1 & 1 \\
1 & 1 & -i\gamma  & 0 \\
0 & 1 & 0 & -i\gamma
\end{array}
\right) ,  \label{eq:H_4x4_open_2}
\end{equation}
exhibits the same antiunitary symmetry based on $\mathbf{U}_{2}$. The two
matrices $\mathbf{H}_{O}$ and $\mathbf{H}_{O2}$ are similar because they are
related as $\mathbf{VH}_{O}\mathbf{V=H}_{O2}$ by means of the unitary matrix
\begin{equation}
\mathbf{V}=\left(
\begin{array}{llll}
1 & 0 & 0 & 0 \\
0 & 0 & 1 & 0 \\
0 & 1 & 0 & 0 \\
0 & 0 & 0 & 1
\end{array}
\right) ,\;\mathbf{V}^{2}=\mathbf{I.}  \label{eq:V}
\end{equation}
For this reason they are isospectral. The matrices (\ref{eq:H_4x4_open}) and
(\ref{eq:H_4x4_open_2}) are associated to the same graphical representation
given in figure~\ref{Fig:Open} with a different labelling of the sites.

Some non-Hermitian operators with tridiagonal matrix representations exhibit
real eigenvalues because they can be transformed into Hermitian operators by
means of suitable transformations\cite{F22}. However, the arguments put
forward in that paper cannot be applied to present matrices that exhibit
complex diagonal elements.

\section{Conclusions}

\label{sec:conclusions}

Throughout this paper it has been our purpose to make two points. First, if
the chosen candidate for the parity operator $\mathcal{P}$ is not a true
parity it is convenient to refer to it simply as a unitary operator and use
the term antiunitary symmetry instead of $\mathcal{PT}$ symmetry. The reason
is that in many cases there are more than one unitary operator $U$
satisfying $U^{2}=\hat{1}$ that may be candidates for generalized $\mathcal{P%
}$ as shown in section~\ref{sec:example} and also in a previous article\cite
{F15}. In this way we avoid fancy names like partial $\mathcal{PT}$ symmetry
or reverse-order partial $\mathcal{PT}$ symmetry\cite{F15}. Second, we show
another example of a non-Hermitian operator, equation (\ref{eq:H_4x4_per}),
that exhibits extremely broken antiunitary symmetry. As in the case of other
models discussed in previous papers\cite{FG14b,FG14c,FG15}, the reason is
the symmetry of the operator in the Hermitian limit ($\gamma =0$ in the
present case) that exhibits degenerate eigenvectors. As argued there,
first-order perturbation theory is a suitable tool for verifying whether the
eigenvalues are complex for small values of the model parameter.

It is worth mentioning recent studies of maximal symmetry breaking in even $%
\mathcal{PT}$-symmetric lattices\cite{JB11} (and references therein)
although such effect does not appear to be related to the point-group
symmetry of the lattice.

\section*{Acknowledgements}

We thank Professor Jacob Barnett for pointing out a misprint in the matrix $%
\mathbf{H}_{O2}$ appearing in an earlier version of this paper.

\end{document}